\documentstyle{EuroPhys}

\def\etal{{\hbox{{\tenit\ et al.\/}\tenrm :\ }}}

\def\And{{\rm and\ }}

\def\stars{\bigskip\centerline{***}\medskip}

\newif\ifboo \boofalse

\input epsf

\begin{document}

\title{Non--monotonous crossover between capillary condensation and interface 
       localisation/delocalisation transition in binary polymer blends} 
\shorttitle{ M. M{\"U}LLER \etal  CAPILLARY CONDENSATION AND INTERFACE LOCALISATION/DELOCALISATION}

\author{ M.\ M\"{u}ller,\inst{1} K.\ Binder,\inst{1} \And E.V. Albano\inst{2} }
\institute{
\inst{1}{Institut f{\"u}r Physik, WA 331, Johannes Gutenberg Universit{\"a}t, D-55099 Mainz, Germany}
\inst{2}{INIFTA, Universidad de la Plata, C.C.\ 16 Suc.\ 4., 1900 La Plata, Argentina}
}
\Date{ 2000}
\euro{}{}{}{}
\rec{December 5, 1999}{April 14, 2000}
\pacs{
\Pacs{05}{70-a}{Phase transitions: general aspects}
\Pacs{68}{45-Gd}{Wetting}
\Pacs{83}{80-Es}{Polymer blends}
}
\maketitle

\begin{abstract}
Within self--consistent field theory we study the phase behaviour of a symmetric binary $AB$ polymer blend confined into a thin
film. The film surfaces interact with the monomers via short range potentials. One surface attracts 
the $A$ component and the corresponding semi--infinite system exhibits a first order wetting transition. 
The surface interaction of the opposite surface is varied as to study the crossover from capillary 
condensation for symmetric surface fields to the interface localisation/delocalisation transition for 
antisymmetric surface fields. In the former case the phase diagram has a single critical point close to 
the bulk critical point.  In the latter case the phase diagram exhibits two 
critical points which correspond to the prewetting critical points of the 
semi--infinite system. The cross\-over between these qualitatively different limiting behaviours occurs 
gradually, however, the critical temperature and the critical composition exhibit a non-monotonic dependence 
on the surface field.
\end{abstract}

\section{Introduction}
We study the phase behaviour of a symmetric binary mixture confined into a slit--like pore. The shift of the critical
point for symmetric film surfaces\cite{NAKANISHI} (capillary condensation) has been studied extensively. The unmixing transition
approaches the critical temperature of the infinite system for thick films. Below the critical temperature the coexisting phases 
differ in their composition at the center of the film. More recently, novel types of phase transitions have been studied
in the case of antisymmetric surface fields,[2-6] {\em i.e.}, one surface attracts the $A$ component 
of a symmetric mixture in exactly 
the same way the opposite surface attracts the $B$ component. Close to the critical temperature in the bulk, enrichment 
layers of the components gradually form at the surfaces and stabilise an interface at the center of the film (soft--mode phase).
It is only close to the temperature\cite{PARRY} of the second order wetting transition in the semi--infinite system that the symmetry 
is spontaneously broken and the interface is located at either surface. Upon increasing the film thickness the temperature of this interface 
localisation/delocalisation transition converges to the wetting temperature rather than the critical temperature 
of the bulk, and the interpretation of the transition in terms of a thin film critical critical point or a wetting transition 
in the limit of large film thickness has been discussed.\cite{TROUBLE} The effect of an external field ({\em i.e.}, gravity)
has been explored.\cite{GRAVITY} Analytical approaches and simulations have considered systems with second order wetting transitions\cite{PARRY,BINDER} 
in the semi--infinite system or the behaviour at bulk coexistence\cite{INDEKEU,TRI} only. This has excluded a  possible interplay between phase behaviour 
and prewetting, which might alter the topology of the phase diagram in thin films,\cite{WET2,WET} from consideration.

In this Letter we revisit the interface localisation/delocalisation transition for a first order wetting transition in the
corresponding semi--infinite system and explore the crossover from the interface localisation/delocalisation transition to  
capillary condensation upon varying the interaction  $\Lambda_2$ at one surface. The interaction $\Lambda_1$ at the other surface 
is kept constant. This is the first systematic theoretical study of boundary conditions which are neither strictly symmetric nor 
antisymmetric. It is clearly pertinent to the interpretation of experiments as the idealized limiting cases are never strictly realized.
Our calculations yield information about the range of asymmetry where the behaviour characteristic of the symmetric and antisymmetric
boundary conditions is observable, and we explore the dependence of the topology of the phase diagrams on the surface interactions. 
Both the phase diagram as a function of the intensive variables incompatibility and chemical potential as well as the binodals are discussed.

\section{Self--consistent field calculations}
We calculate the phase behaviour of a confined $AB$ mixture within the self-consistent field theory of  Gaussian 
polymers.[12-14]
Polymer mixtures in confined geometries are interesting for many applications ({\em e.g.}, coating, lubrication), and consequentially 
have attracted abiding recent 
attention.\cite{REVS} In these systems the soft mode phase mentioned above has been experimentally studied.\cite{KERLE} Thus we
focus on these systems, although we believe that our results carry over to other confined mixtures at least qualitatively.
We consider a film with volume $V_0=\Delta_0 \times L \times L$. 
$\Delta_0$ denotes the film thickness, while $L$ is the lateral extension of the film. The density at the film surfaces decreases 
to zero in a boundary region of width $\Delta_w$ according to\cite{FILM} 
\begin{equation}
\Phi_0(x) = \left\{ \begin{array}{ll}
               \frac{1-\cos\left( \frac{\pi x}{\Delta_w}\right)}{2} 
	&	\mbox{{\small ; $0 \leq x \leq \Delta_w$}} \\
	       1                                                   
		& \mbox{{\small ;  $\Delta_w \leq x \leq \Delta_0 - \Delta_w$}} \\
               \frac{1-\cos\left( \frac{\pi (\Delta_0 - x)}{\Delta_w}\right)}{2}  
		 & \mbox{{\small ; $\Delta_0 - \Delta_w \leq x \leq \Delta_0$}}
             \end{array}\right. 
	     \label{eqn:dens}
\end{equation}
where $\Phi_0$ denotes the ratio of the monomer density and the value $\rho$ in the middle of the film.
The thickness $\Delta$ of an equivalent film with constant monomer density $\rho$ is 
$\Delta=\Delta_0-\Delta_w$. We choose $\Delta_w=0.15 R_e$,\cite{FILM} where $R_e$ is the end--to--end distance.
Both surfaces interact with the monomer species via a short range potential $H$:
\begin{equation}
H(x) = \left\{ \begin{array}{ll}
            \frac{4 \Lambda_1 R_e\left\{1+\cos\left( \frac{\pi x}{\Delta_w}\right)\right\}}{\Delta_w} 
	    & \mbox{{\small ; $0\leq x \leq \Delta_w$ }} \\
            0                                                                                         
	    & \mbox{{\small ;  $\Delta_w \leq x \leq \Delta_0 - \Delta_w$}} \\
            \frac{4 \Lambda_2 R_e\left\{1+\cos\left( \frac{\pi (\Delta_0 - x)}{\Delta_w}\right)\right\}}{\Delta_w}
	    & \mbox{{\small ; $\Delta_0 - \Delta_w \leq x \leq \Delta_0$}}
        \end{array}\right.
\end{equation}
$H>0$ is attractive for the $A$ monomers and repulsive for the $B$ species.
The normalisation of the surface fields $\Lambda_1$ and $\Lambda_2$, which act on the monomers close to the 
left and the right surface, is chosen such that the integrated interaction energy between the surface and the 
monomers is independent of the width of the boundary region $\Delta_w$.

$A$ and $B$ polymers contain $N$ monomers and are structurally symmetric.
The polymer conformations $\{ {\bf r}_\alpha(\tau)\}$ determine the microscopic $A$ monomer density 
$\hat \Phi_A({\bf r}) = \frac{N}{\rho} \sum_{\alpha=0}^{n_A} \int_0^1 {\rm d}\tau \times$\\
$ \delta\left({\bf r}-{\bf r}_\alpha(\tau)\right)$,
where the sum runs over all $n_A$ $A$ polymers in the system and $0 \leq \tau \leq 1$ parameterises the 
contour of the Gaussian polymer. A similar expression holds for $\hat \Phi_B({\bf r})$.
With this definition the semi--grandcanonical partition function takes the form:
\begin{eqnarray}
{\cal Z} &\sim \sum_{n_A=1}^n \frac{e^{+\Delta \mu n_A/2k_BT}}{n_A!} \;\;
                              \frac{e^{-\Delta \mu n_B/2k_BT}}{n_B!} & 
               \int {\cal D}_A[{\bf r}]  {\cal P}_A[{\bf r}] 
               \int {\cal D}_B[{\bf r}]  {\cal P}_B[{\bf r}]  \;\delta\left( \Phi_0 - \hat \Phi_A - \hat \Phi_B \right) \nonumber \\
         &&       \;\times \;\exp\left( - \rho \int {\rm d}^3{\bf r} 
                \left\{ \chi \hat \Phi_A \hat \Phi_B - H(\hat \Phi_A-\hat \Phi_B \right\}\right)   
\end{eqnarray}
where $n=n_A+n_B$ and $\Delta \mu$ represents the exchange potential between $A$ and $B$ polymers.
The functional integral ${\cal D}$ sums over all conformations of the polymers, and 
${\cal P}[{\bf r}] \sim \exp \left(- \frac{3}{2R_e^2} \int_0^1 {\rm d}\tau \;\left(\frac{{\rm d}{\bf r}}{{\rm d}\tau}\right)^2 \right)$
denotes the statistical weight of a non--interacting Gaussian polymer.  The second factor 
enforces the monomer density profile to comply with Eq.(\ref{eqn:dens}) (incompressibility). 
The Boltzmann factor in the partition function incorporates the 
thermal repulsion between unlike monomers and the interactions between  monomers and surfaces. 
The strength of the repulsion is described by the Flory--Huggins parameter $\chi$.

In mean field approximation the free energy is obtained as the extremum of the functional:
\begin{eqnarray}
\frac{{\cal G}[W_A,W_B,\Phi_A,\Phi_B,\Xi]}{n k_BT}  \equiv  +\; \ln\frac{n}{V_0}  - \;\ln {\cal Q}           
     + \; \frac{1}{V} \int {\rm d}^3{\bf r} \;\; \chi N \Phi_A \Phi_B                      
                                              -  H N \left\{ \Phi_A-\Phi_B\right\}\nonumber \\
     - \; \frac{1}{V} \int {\rm d}^3{\bf r} \;\;  \left\{ W_A \Phi_A + W_B \Phi_B\right\} +  \Xi \left\{\Phi_0 -  \Phi_A -  \Phi_B\right\} 
						 \label{eqn:F}
\end{eqnarray}
with respect to its five arguments. ${\cal Q}_A$ denotes the single chain partition function:
\begin{equation}
{\cal Q}_A[W_A] = \frac{1}{V_0} \int {\cal D}_1[{\bf r}] {\cal P}_1[{\bf r}] \;\; e^{ - \int_0^1 {\rm d}\tau\; W_A({\bf r}(\tau)) }
\end{equation}
a similar expression holds for ${\cal Q}_B$, and
${\cal Q} =  \exp(\Delta \mu/2k_BT) {\cal Q}_A + \exp(-\Delta \mu/2k_BT) {\cal Q}_B$.

The values of $W_A,W_B,\Phi_A,\Phi_B,\Xi$ which extremize the free energy functional are denoted by lower--case letters 
and satisfy the self-consistent set of equations
\begin{equation}
w_A({\bf r}) = \chi N \phi_B({\bf r}) - H({\bf r}) N + \xi({\bf r}) \qquad \mbox{and} \qquad
\phi_A({\bf r}) = -\frac{V}{{\cal Q}} \frac{{\cal DQ}_A}{{\cal D}w_A({\bf r})} 
\label{eqn:scf}
\end{equation}
$\Phi_0({\bf r}) = \phi_A({\bf r}) + \phi_B({\bf r})$, and similar expressions for $\phi_B$ and $w_B$.

To calculate the monomer density we employ the end segment distribution $q_A({\bf r},t)$
\begin{equation}
q_A({\bf r},\tau) = \int_0^\tau {\cal D}_1[{\bf r}] {\cal P}_1[{\bf r}] \delta({\bf r} - {\bf r} (\tau)) 
e^{- \int_0^\tau {\rm d}t\; w_a({\bf r}(t))}
\end{equation}
It satisfies the diffusion equation:
\begin{equation}
\frac{\partial q_A({\bf r},\tau)}{\partial \tau} = \frac{R_e^2}{6} \triangle q_A({\bf r},\tau) - w_A({\bf r}) q_A({\bf r},\tau)
\end{equation}
Then, the $A$ monomer density and the single chain partition  can be calculated via
\begin{equation}
\phi_A({\bf r}) = \frac{V e^{\Delta \mu/2k_BT}}{V_0 {\cal Q}}\int_0^1 {\rm d}\tau\; q_A({\bf r},\tau) q_A({\bf r},1-\tau)
\qquad \mbox{and} \qquad
{\cal Q}_A = \frac{1}{V_0} \int {\rm d}^3{\bf r}\; q_A({\bf r} ,1)
\end{equation}
We expand the spatial dependence of the densities and fields in a set of orthonormal 
functions $\{f_k(x) = \sqrt{2} \sin(\pi k x/\Delta_0) \}$ with $k=1,2,\cdots$.\cite{MARK,FILM} This procedure results in a set of 
non--linear equations which are solved by a Newton--Raphson--like method. We use up to 
80 basis functions and achieve a relative accuracy $10^{-4}$ in the free energy. For symmetric boundary fields 
only components with an odd index $k$ are
non--zero. Substituting the extremal values of the densities and fields into the free energy functional (\ref{eqn:F}) we 
calculate the free energy $G$ of the different phases. At coexistence the two phases have equal semi--grandcanonical 
potential.

\section{Results}
\begin{figure}[htbp]
    \begin{minipage}[t]{85mm}%
       \mbox{
        ({\bf a})
        \setlength{\epsfxsize}{7.8cm}
        \epsffile{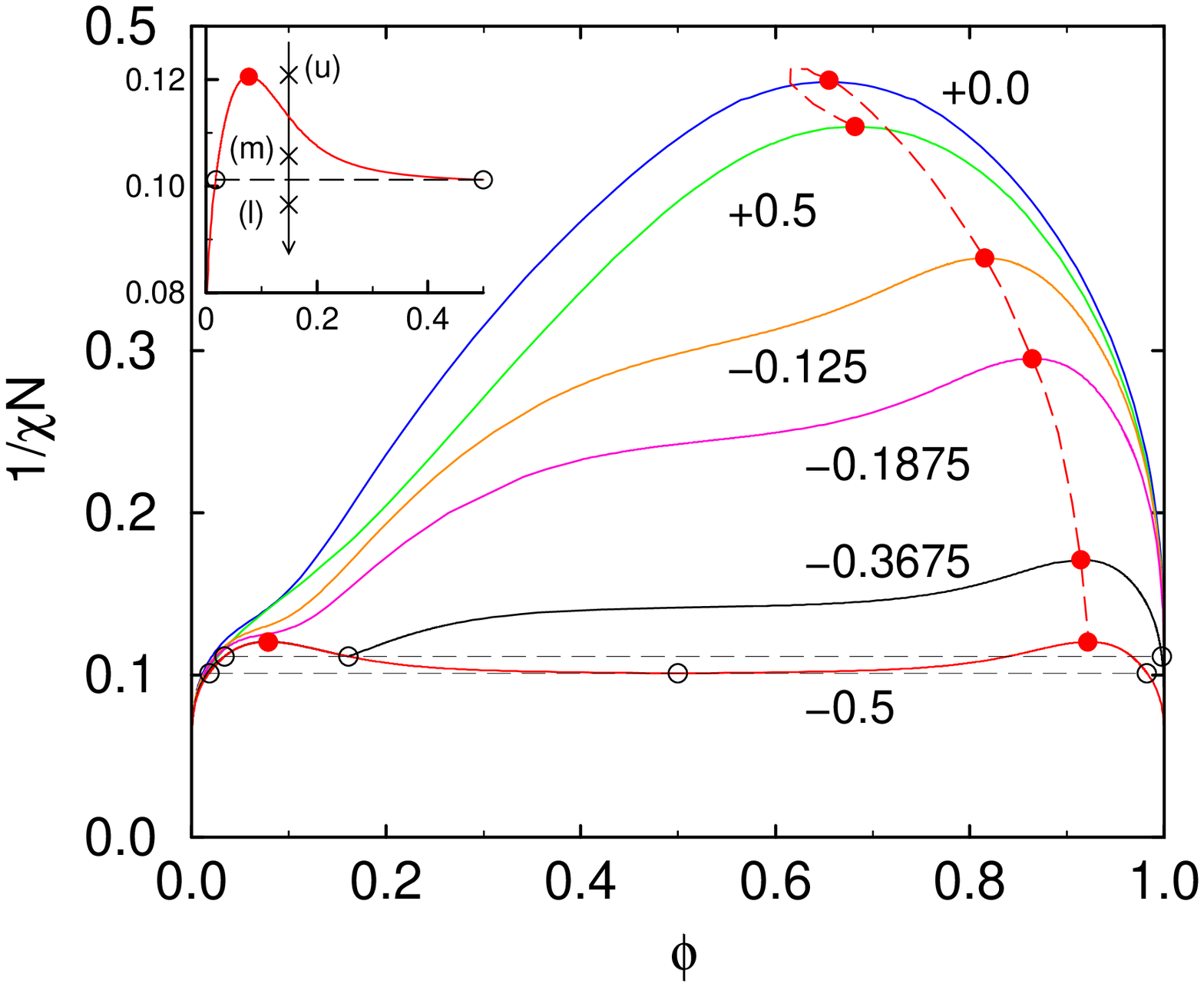}
       }
    \end{minipage}
    \hspace*{-1cm}
    \begin{minipage}[t]{50mm}%
       \mbox{
        ({\bf b})
        \setlength{\epsfxsize}{4.5cm}
        \epsffile{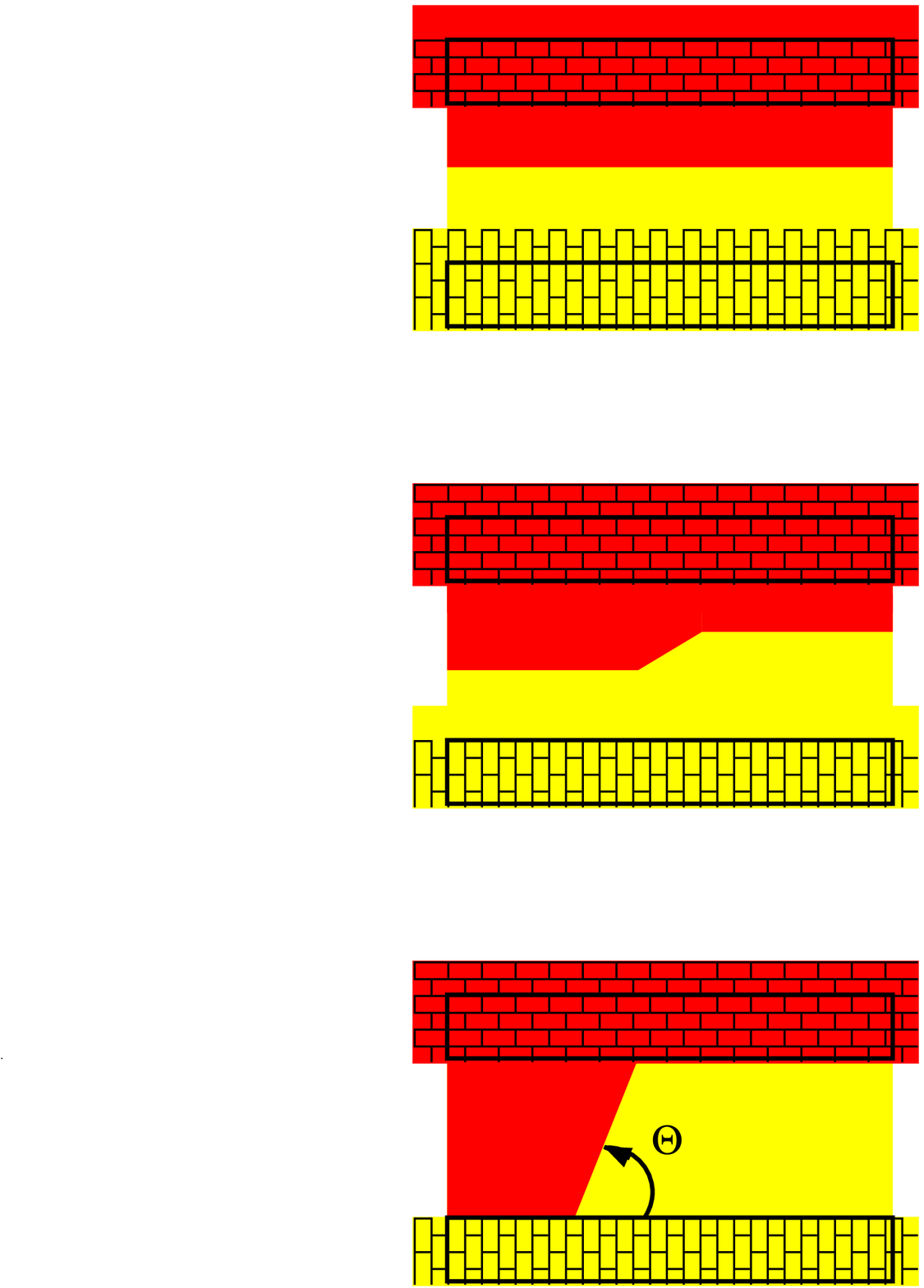}
	}
    \end{minipage}%
    \hspace*{-2.15cm}
    \begin{minipage}[b]{23mm}%
      {
      $(u)$

      \vspace*{2cm}

      $(m)$

      \vspace*{2cm}

      $(l)$
      \vspace*{1.5cm}
      }
    \end{minipage}%
    \\
    \begin{minipage}[b]{140mm}%
       \caption{
       \label{fig:1} ({\bf a}) Binodals for $\Delta_0=2.6 R_e$ and $\Lambda_1 N=0.5$. $\Lambda_2 N$ varies as 
       indicated in the key.  The dashed curve shows the location of the critical points. Filled circles 
       mark critical points, open circles/dashed horizontal lines denote three phase coexistence for 
       $\Lambda_2 N=-0.3675$ and $-0.5$. The inset presents part of the phase boundary for antisymmetric 
       boundaries. 
       ({\bf b}) Schematic temperature dependence for antisymmetric boundaries. The three profiles
       correspond to the situations (u),(m), and (l) in the inset of ({\bf a}). 
       }
    \end{minipage}%
\end{figure}

The phase boundaries as a function of the surface fields are presented in Fig.\ref{fig:1}. First, we consider the antisymmetric boundary 
condition  $\Lambda_1N=-\Lambda_2N=0.5$. One half of the phase diagram is enlarged in the inset. If we reduce the temperature along the 
symmetry axis of the phase diagram $\phi=1/2$, an
interface is stabilized at the center of the film for $\chi N <\chi_{\rm triple} N$, and at $\chi_{\rm triple} N$ we encounter
a first order interface localisation/delocalisation transition at which the interface jumps discontinuously to one of the two surfaces.
At this triple point phases with compositions $\phi_{\rm triple}$, $1/2$ and $1-\phi_{\rm triple}$ coexist. 
This behaviour for $\Delta\mu=0$ has been confirmed by computer simulations of Ising models.\cite{BINDER,TRI}
For all other compositions in the range
$[\phi_{\rm triple}:1-\phi_{\rm triple}]$, however, one encounters two transitions upon cooling. Consider cooling at $\phi_{\rm triple}<\phi<1/2$ as indicated by
the arrow in the inset. At high temperatures enrichment layers gradually form at the surfaces. Slightly below the critical temperature of the bulk an interface 
is stabilized which separates a thin $A$--rich layer at one surface from a thicker $B$--rich layer at the opposite surface (cf.\ Fig.\ref{fig:1}{\bf b} {\bf u}pper 
panel). Laterally, the system is homogenous (soft--mode phase) and an $AB$ interface runs parallel to the surfaces.
Upon cooling, we first encounter a phase separation into lateral regions with a thin and a thick $A$--rich layer(cf.\ Fig.\ref{fig:1}{\bf b} {\bf m}iddle panel). 
This phase coexistence is the analogy of the prewetting line in a semi--infinite system. Since the coexistence involves only finite layer thicknesses 
it also persists in a thin film, provided the film thickness is sufficiently large. The different layer thicknesses correspond to different compositions of
the film and give rise to two symmetric coexistence regions. This prediction has not yet been observed in computer simulations\cite{BINDER,TRI} or 
experiments.\cite{KERLE,JONES} 
Upon further cooling, we encounter a second transition at $\chi_{\rm triple} N$ where the two phase coexistences merge and the system laterally segregates 
into $A$--rich and $B$--rich regions(cf.\ Fig.\ref{fig:1}{\bf b} {\bf l}ower panel). For temperatures far below the triple temperature 
the interface between the coexisting phases runs straight across the film. The angle between the interface and the surface
corresponds to the contact angle $\Theta$ of droplets in the semi--infinite system. Upon increasing the film thickness the triple temperature approaches the
wetting transition temperature, the two coexistence regions become narrower and converge to the prewetting lines, and the critical temperature of the film 
tends to the prewetting critical temperature of the semi--infinite system.\cite{PHYSA}
\begin{figure}[htbp]
    \begin{minipage}[t]{85mm}%
       \mbox{
        \setlength{\epsfxsize}{7.8cm}
        \epsffile{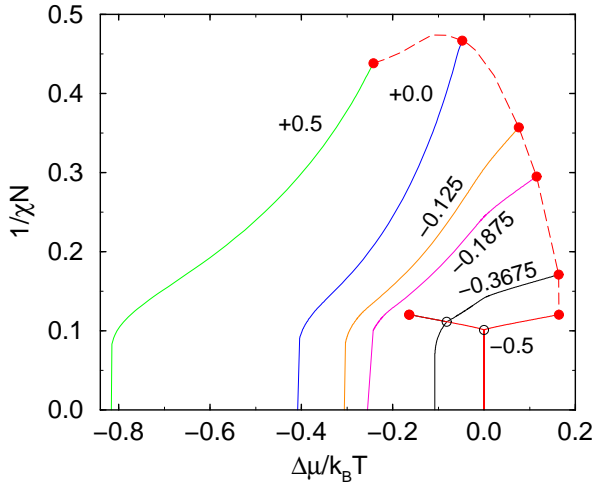}
       }
    \end{minipage}
    \hspace*{0.5cm}
    \begin{minipage}[b]{50mm}%
       \caption{
       \label{fig:2} Coexistence curves in the $\chi N$--$\Delta\mu$ plane. $\Lambda_2 N$ varies
       according to the key. Parameters and symbols as in Fig.\protect\ref{fig:1}. The ``quasi--prewetting'' 
       lines for $\Delta \mu<0$ and $\Lambda_2N=-0.3675$ and $-0.5$ are indistinguishable, because they are
       associated with the prewetting behaviour of the surface with interaction $\Lambda_1N=+0.5$.
       }
    \end{minipage}%
\end{figure}
The phase coexistence in terms of incompatibility $\chi N$ and exchange potential $\Delta \mu$ is shown in Fig.\ref{fig:2}.
Although the Hamiltonian of the system is symmetric with respect to the exchange $A \rightleftharpoons B$, phase coexistence occurs at $\Delta \mu=0$ 
only below the triple temperature. At $\chi_{\rm triple} N$ the coexistence curve bifurcates and the branches move away from the symmetry axis. The two 
coexistence lines resemble the prewetting lines which correspond to the first order wetting transitions at the two surfaces. 

At low temperatures $\chi\gg\chi_{\rm triple}$ the two coexisting phases are almost pure and the coexistence value of the chemical potential is 
given by $\Delta \mu_{\rm coex}/k_BT=-4(\Lambda_1+\Lambda_2)NR_e/\Delta$. Upon increasing 
$\Lambda_2$ this value shifts to more negative values. Only for large film thickness the two phase region on the $B$--rich side of the phase 
diagram remains.\cite{WET2,WET} For thin films, however, $\Delta \mu_{\rm coex}$ is shifted to negative values which are eventually smaller than the chemical 
potential of the prewetting critical point at the surface which attracts $A$ (cf.\ Fig.\ref{fig:2}). For the parameters chosen this occurs roughly around $\Lambda_2 N \approx -0.2$. In this case the two phase region on the $B$--rich side as well as the associated critical point
disappears upon increasing $\Lambda_2$ further.

The critical point on the $A$--rich side of the phase diagram shifts to higher temperatures upon increasing $\Lambda_2$ similar to the prewetting 
at the corresponding surface. Note that $\chi_{\rm wet}N \approx 24 (\Lambda_2 N)^2$ for strong segregation.\cite{WET} 
Although the film surfaces favor (on average) the $A$ component of the mixture, the critical point occurs at $\Delta \mu>0$ for $\Lambda_2 N \stackrel{<}{~}-0.25$.
Around $\phi=1/2$ or $\Delta \mu_{\rm coex}=0$ the $B$--rich binodal shows a concave curvature, which is the remanent of the wetting transition at the surface which favours $B$. 
Upon increasing $\Lambda_2$ to positive values this feature disappears, of course. 

\begin{figure}[htbp]
    \begin{minipage}[t]{85mm}%
       \mbox{
        \setlength{\epsfxsize}{7.8cm}
        \epsffile{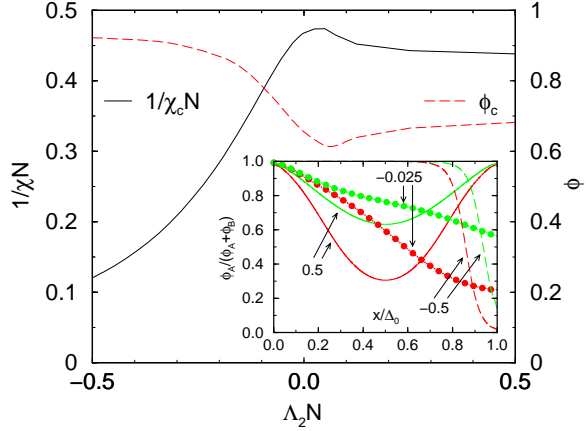}
       }
    \end{minipage}
    \hspace*{0.5cm}
    \begin{minipage}[b]{50mm}%
       \caption{
       \label{fig:3} Dependence of the critical temperature (left axis) and density (right axis) on the surface field $\Lambda N_2$. 
       $\Lambda_1N=0.5$ and $\Delta_0=2.6 R_e$.
       The inset presents the composition profiles of the coexisting phases $3\%$ below the critical temperature for $\Lambda_2 N= 0.5$ 
       (solid line), $-0.025$ (dotted line) and $-0.5$ (dashed line).
       }
    \end{minipage}%
\end{figure}
For an almost neutral surface $\Lambda_2 N \approx 0$ the critical temperature 
passes through a maximum and the critical temperature of the capillary condensation is approached from above.\cite{COMMENT1} The surface field dependence of the
critical temperature and critical density is investigated in Fig.\ref{fig:3}.
This non-monotonic behaviour of the critical temperature and density can be rationalised as follows:
For $\Lambda_2 = \Lambda_1$ the composition profiles of the two coexisting phases close to the critical point are symmetric (c.f.\ Fig.\ref{fig:3} inset). 
It is the difference in the composition at the center which distinguishes between the phases and  vanishes at the critical point. Decreasing $\Lambda_2$ we reduce
the influence of the surfaces, and the critical temperature increases and the critical composition decreases, respectively, as both tend towards their 
bulk values.\cite{COMMENT3}
For large negative values of $\Lambda_2$ the transition is qualitatively different. It is associated with the prewetting transition, {\em i.e.},
the profiles of the coexisting phases resemble a profile across an interface and it is the location of the interface which distinguishes
the two phases (c.f.\ Fig.\ref{fig:3} inset). Decreasing $\Lambda_2$ shifts the wetting transition to lower temperatures and higher $A$ content.
The gradual crossover between the two distinct behaviours occurs around $\Lambda_2 \approx 0$.\cite{COMMENT} 
The corresponding profiles are included into Fig.\ref{fig:3} (inset).
In both phases the composition at the surface favouring $A$ is almost saturated and the composition decreases to a value larger or smaller than 1/2 at the
almost neutral surface. Upon approaching the critical point the composition difference at the neutral surface vanishes. This is similar to the capillary condensation
mechanism for $\Lambda_2>0$. Both profiles, however, can also be perceived as very broad interfaces (with a width comparable to the film thickness) with 
different centers. This resembles the prewetting--like behaviour for $\Lambda_2<0$.

\section{Discussion}
In summary, we have explored the unusual dependence of the phase behaviour of a confined symmetric mixture on the surface fields. For nearly antisymmetric
surface fields which are strong enough to produce a first order wetting transition our mean field calculations predict two coexistence regions which correspond 
to the prewetting transition at either surface. We vary the surface interactions  from  antisymmetric to symmetric and
the critical temperature(composition) passes through a maximum(minimum) when one surface is almost neutral. 
We expect the qualitative dependence of the phase diagram on the surface fields to be generic. Fluctuations, which are not included in the mean field
calculations, impart 2D Ising rather than  mean field behaviour to the critical points. Their importance is, however, restricted to a narrow region around 
the critical point in the limit of strong interpenetration $\rho R_e^3/N \gg 1$. Capillary waves increase the range of the 
effective interaction\cite{PARRY2} between interface and surface by a factor $1+\omega/2$ where $\omega=k_BT/4\pi\xi^2\sigma$. 
$\xi$ and $\sigma$ denote the correlation length and interfacial tension between the coexisting bulk phases, respectively. Simulations\cite{WET} 
indicate that $1/\omega \sim \sqrt{\chi N} f(\chi N) \rho R_e^3/N$, where the scaling function $f(\chi N)$ approaches 
a constant value for $\chi N \to \infty$. Since the wetting transition in binary polymer blends is typically first order and occurs
at strong segregation,\cite{WET} the capillary parameter $\omega$ is small.
Experiments\cite{KERLE,JONES} on 
polymer blends have explored the phase behaviour in confined geometry, and these systems as well as computer simulations might prove convenient for testing 
our predictions.

We are currently exploring the thickness dependence of the phase diagram for antisymmetric surfaces within the 
self--consistent field theory and a Landau--Ginzburg approach.\cite{PHYSA} Contrary to the behaviour close to a second order wetting transition, we find
that the triple point approaches the temperature of the first order wetting transition from above upon increasing the film thickness. The transition between 
first and second order interface localisation/delocalisation transition and the behaviour at the tricritical point will be elucidated.

\stars
It is a pleasure to thank E.\ Reister and F.\ Schmid for helpful discussions. Financial support was provided by the DFG
under grant Bi314/17 within the Priority Program ``Wetting and Structure Formation at Interfaces'', the VW Stiftung, and PROALAR2000.


\begin{thebibliography}{99}
\bibitem{COMMENT1}  More general, the shift in the critical temperature can be described by a finite--size scaling ansatz of the form:
		    $\chi_c(\Delta)/\chi_c({\rm bulk})-1 = \Delta^{-1/\nu} f(\Delta^{\Delta_1/\nu} \Lambda_1N,\Delta^{\Delta_1/\nu} \Lambda_2N)$ and
		    where $\nu =1/2$ and $\Delta_1=1$ denote the bulk correlation length and  the 
		    surface gap exponent in mean field theory.\protect{\cite{NAKANISHI}} Our calculations indicate a non--monotonic 
		    behaviour of the scaling function $f$.
\bibitem{COMMENT3}  The critical point for completely neutral surfaces $\Lambda_1 = \Lambda_2=0$ occurs at
                    $\phi_c=1/2$ and $1/\chi_c N = 0.493$, which is very close to the bulk critical point $\chi_cN=2$.
\end{thebibliography}
\end{document}